# Andreev reflection and bound state formation in a ballistic two-dimensional electron gas probed by a quantum point contact

Hiroshi Irie[1,*], Clemens Todt[1], Norio Kumada[1], Yuichi Harada[1,†], Hiroki Sugiyama[2], Tatsushi Akazaki[1,‡] and Koji Muraki[1]

[1] *NTT Basic Research Laboratories, NTT Corporation, 3-1 Morinosato-Wakamiya, Atsugi, Kanagawa 243-0198, Japan*
[2] *NTT Device Technology Laboratories, NTT Corporation, 3-1 Morinosato-Wakamiya, Atsugi, Kanagawa 243-0198, Japan*

We study coherent transport and bound-state formation of Bogoliubov quasiparticles in a high-mobility $In_{0.75}Ga_{0.25}As$ two-dimensional electron gas (2DEG) coupled to a superconducting Nb electrode by means of a quantum point contact (QPC) as a tunable single-mode probe. Below the superconducting critical temperature of Nb, the QPC shows a single-channel conductance greater than the conductance quantum $2e^2/h$ at zero bias, which indicates the presence of Andreev-reflected quasiparticles, time-reversed states of the injected electron, returning back through the QPC. The marked sensitivity of the conductance enhancement to voltage bias and perpendicular magnetic field suggests a mechanism analogous to reflectionless tunneling—a hallmark of phase-coherent transport, with the boundary of the 2DEG cavity playing the role of scatters. When the QPC transmission is reduced to the tunneling regime, the differential conductance vs bias voltage probes the single-particle density of states in the proximity area. Measured conductance spectra show a double peak within the superconducting gap of Nb, demonstrating the formation of Andreev bound states in the 2DEG. Both of these results, obtained in the open and closed geometries, underpin the coherent nature of quasiparticles, i.e., phase-coherent Andreev reflection at the InGaAs/Nb interface and coherent propagation in the ballistic 2DEG.

## I. Introduction

The superconducting proximity effect in superconductor-normal metal (SN) hybrid structures has gained increased interest for both studying exotic quantum phases [1-7] and developing novel electronic devices [8-11]. In such hybrid structures, charge transport near the SN interface is governed by quasiparticles generated by phase-coherent Andreev reflections (ARs) at the SN interface [12-14]. Andreev-reflected quasiparticles, being charge- and time-reversed states of those impinging on the SN interface, give rise to unique transport properties such as conductance doubling and retroreflection, which respectively have been demonstrated using point-contact [15] and magneto-focusing [16] spectroscopy in the ballistic regime. In the diffusive regime, on the other hand, the retroreflection property leads to reflectionless tunneling, observed as a zero-bias conductance peak [17]. As also manifested in the reflectionless tunneling, Andreev-reflected quasiparticles carry information about the macroscopic phase of the superconductor by storing it in their dynamical phase, thereby bringing superconducting correlation into the N region.

When the N region is sufficiently small compared to the coherence length and the mean free path, quasiparticles are confined to form (quasi)bound states known as Andreev bound states (ABSs) [18,19]. ABSs can form in both SN and SNS junctions. While it is theoretically well-established that superconducting Josephson current in SNS junctions is mediated by ABSs [19,20], it is only recently that direct observation of ABSs by tunneling and microwave spectroscopy [5,21-29] has become possible. However, the short mean free path in the N region has limited these studies to systems with the size of the N region comparable to or smaller than the Fermi wavelength $\lambda_f$. On the other hand, individual processes of AR, which would be responsible for ABS formation in a confined geometry, have only been studied in open geometries, leaving experiments that bridge between the two regimes unexplored.

In this paper, we study an SN junction consisting of a superconducting Nb electrode and the high-mobility $In_{0.75}Ga_{0.25}As$ two-dimensional electron gas (2DEG). By taking advantage of a long mean free path of the 2DEG, we can explore the quasiparticle transport in the ballistic regime. By utilizing a quantum point contact (QPC) formed in the vicinity of the SN interface, we can study the effects of the boundary condition on the quasiparticle transport by tuning the transmission probability from unity to zero. With unity transmission, the Andreev-reflected quasiparticles, which trace back the path of the incoming electrons, transmit through the QPC. By comparing the single-channel conductance with the conductance quantum $2e^2/h$, which is expected for a QPC with normal contacts [30,31], we are able to detect the transmission of the returning quasiparticles. When tuned in the low transmission regime, the QPC works both as a confining potential defining ABSs and a tunneling barrier for the spectroscopy of the ABSs. In the following sections, we present data on the QPC conductance under two boundary conditions, i.e., full and near-zero transmission, with which we demonstrate the phase-coherent nature of the Andreev-reflected quasiparticles in the ballistic regime.

## II. Experiments

---
[*] irie.hiroshi@lab.ntt.co.jp
[†] Present address: Art, Science and Technology Centre for Cooperative Research, Kyushu University, 6-1 Kasuga-koen Kasuga, Fukuoka 816-8580, Japan.
[‡] Present address: Department of Electrical Engineering and Information Science, National Institute of Technology Kochi College, 200-1 Monobe Otsu, Nankoku, Kochi 783-8508, Japan

Figure 1(a) schematically shows the device structure of our SN hybrid QPC. The InGaAs 2DEG, which serves as a ballistic N, is bounded by an interface with Nb on one side and a QPC on the other side, which together form a ballistic cavity. The device can thus be viewed as an NINS structure, where the QPC plays the role of the insulator (I) with tunable transmission $T_{QPC}$. The energy band diagrams for the open-channel ($T_{QPC} \approx 1$) and tunneling ($T_{QPC} \ll 1$) regimes are shown in Figs. 1(b) and 1(c), respectively. At the SN interface, electron- and hole-like quasiparticles in the N region are transformed into each other by an AR process. Note that the AR changes not only the charge (electron ↔ hole) but also the kinetic energy of the quasiparticles ($\varepsilon$ for electron ↔ $-\varepsilon$ for hole, where $\varepsilon$ represents the energy with respect to the Fermi level). In the open-channel regime [Fig. 1(b)], the Andreev-reflected hole returns to the left reservoir, which results in a doubling of the QPC conductance. In the tunneling regime [Fig. 1(c)], quasiparticles are reflected at the NI interface and are confined in the ballistic cavity. If the quasiparticles preserve their phase coherence during successive reflections, ABSs emerge as resonance levels within the superconducting gap. Note that, as a result of particle-hole symmetry, ABSs always come in pairs with energy levels symmetric with respect to the Fermi level.

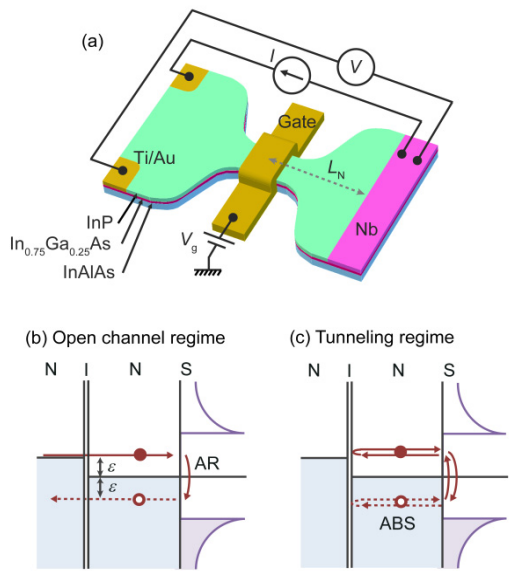

FIG. 1. (color online) Schematic drawings of (a) the SN hybrid QPC studied and the energy band diagrams for (b) the open-channel regime and (c) the tunneling regime. In (b) and (c), the electron- and hole-like quasiparticle are shown by solid and open circles.

As illustrated in Fig. 1(a), a Nb electrode is coupled with the 2DEG confined in an InGaAs/InAlAs/InP heterostructure whose layer structure from the bottom to the surface comprises a semi-insulating InP substrate, a 200-nm-thick $In_{0.52}Al_{0.48}As$ buffer, a 6-nm-thick Si-doped ($4 \times 10^{18}$ cm$^{-3}$) $In_{0.52}Al_{0.48}As$, a 10-nm-thick $In_{0.52}Al_{0.48}As$, a 2DEG layer consisting of $In_{0.53}Ga_{0.47}As/In_{0.75}Ga_{0.25}As/In_{0.53}Ga_{0.47}As$ (2.5/8/5 nm), a 3-nm-thick $In_{0.52}Al_{0.48}As$, and a 5-nm-thick InP cap [32]. The QPC is fabricated by etching the heterostructure into a narrow constriction (120-nm wide and 200-nm long), followed by atomic layer deposition of a 20-nm-thick $Al_2O_3$ insulator and e-beam evaporation of a 80-nm-wide Ti/Au (10/70-nm thick) wrap gate [32]. To fabricate a 2DEG/Nb interface with a low barrier height, the top InP and upper InAlAs layers in the contact region were removed by selective wet etching of InP and subsequent *in-situ* Ar plasma etching in the same chamber as that for the Nb deposition. The thickness of Nb was chosen to be 80 nm, which is larger than the London penetration depth (~40 nm). The distance $L_N$ between the SN interface and the center of the QPC is 220 nm. Two separate Ti/Au ohmic contacts to the 2DEG were made with the same technique as that for the Nb contact. These ohmic contacts are located at a much greater distance of ~100 μm from the QPC to prevent the normal reflection (NR) at the interface from influencing on the QPC conductance. More details about the device fabrication can be found in Ref. [33].

The heterostructure wafer we used hosts a 2DEG with electron density $n_s = 1.9 \times 10^{12}$ cm$^{-2}$ and mobility $\mu_e = 156{,}000$ cm$^2$/Vs, as determined from magnetotransport measurements at 1.8 K on a Hall bar device simultaneously fabricated on the same chip. The corresponding elastic mean free path $l_e$ ($= \hbar\mu_e\sqrt{2\pi n_s}/e$) of 3.5 μm is an order of magnitude longer than $L_N$ (= 220 nm), which places the system in the ballistic regime. The Nb's superconducting gap $\Delta_0 = 1.28$ meV, which is deduced from the measured superconducting transition temperature $T_c = 8.4$ K, translates into the coherence length $\xi_0 = 152$ nm of the 2DEG according to the relation $\xi_0 = \hbar v_{fN}/\pi\Delta_0$, where $v_{fN}$ is the Fermi velocity of the 2DEG. Here we used $v_{fN} = 9.3 \times 10^5$ m/s ($= \hbar\sqrt{2\pi n_s}/m^*$), which was obtained from $n_s$ and the effective mass $m^* = 0.043\ m_e$ ($m_e$ is the electron rest mass) that was estimated from the temperature dependence of the Shubnikov-de Haas oscillations. Comparing $\xi_0$ with the system size $L_N$ suggests that the proximity effect affects the entire N region between the SN interface and the QPC.

Transport measurements were performed using a lock-in technique at 71.3 Hz in a quasi-four-point configuration, where two Au wires are separately attached to the Nb (and the Ti/Au ohmic electrodes) as current and voltage leads [see Fig. 1(a)]. To study the bias dependence, a dc voltage $V_{dc}$ was superimposed on the ac lock-in excitation using a transformer. All measurements presented hereafter were carried out in a $^3$He refrigerator at temperatures ranging from 240 mK to 10 K.

### III. Conductance enhancement via AR in the open-channel regime

To investigate ballistic transport of Andreev-reflected quasiparticles, we first examine the effects of AR on the QPC conductance in the open-channel regime. Figure 2(a) compares the zero-bias differential conductance $dI/dV$ measured at $T = 240$ mK and 10 K, plotted as a function of gate voltage $V_g$. At $T = 10$ K ($> T_c$), $dI/dV$ exhibits conductance quantization in units of $2e^2/h$, as expected for a QPC with normal contacts [30,31]. The $dI/dV$ values of the plateaus are slightly below the multiples of $2e^2/h$. This deviation can be explained by assuming a series resistance of $R_c = 230\ \Omega$, which we ascribe to the contact resistance at the 2DEG/Nb interface.

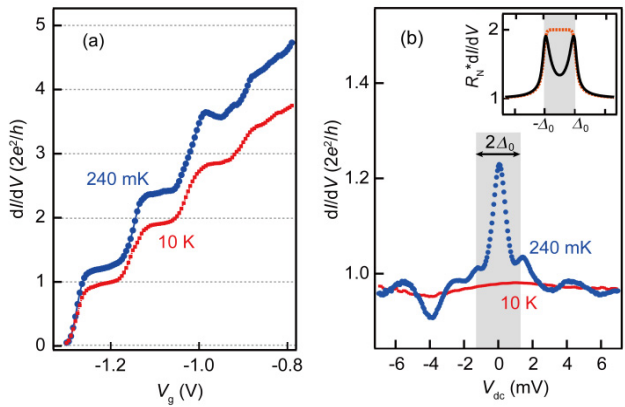

FIG. 2. (color online)  (a) $dI/dV$ at zero bias vs $V_g$ at $T = 240$ mK and 10 K. (b) $dI/dV$ vs $V_{dc}$ at $T = 240$ mK and 10 K. The shaded region represents a bias range within the superconducting gap of Nb. The inset shows simulated $dI/dV$ spectra with the BTK model for SN interfaces with (solid) and without (broken) a potential barrier. A dimensionless barrier height $Z$ of 0.4 [12] is assumed in the calculation.

At $T = 240$ mK ($< T_c$), $dI/dV$ also shows a stepwise change, but with the step heights increased to $1.25 \times 2e^2/h$. The increased step heights indicate that the conductance of each transport mode surpasses the conductance quantum, which arises because transmission of one electron through the QPC is followed by the return of an Andreev-reflected hole back through the QPC. Thus, the observed conductance enhancement is evidence that the proximity effect from the SN interface extends to the QPC region. Note that, in the open-channel regime, AR occurs only once in our SN junction, since all the Andreev reflected holes in the relevant mode are transmitted backwards through the QPC owing to their retroreflection property. Consequently, a maximum conductance of $2 \times 2e^2/h$ is expected for a perfect AR in SN junctions [12,13]. The reduced enhancement factor observed in our SN junction is due to finite NRs coexisting with ARs.

It is worth mentioning the difference between the SN junctions studied here and SNS Josephson junctions studied previously [33-36], in which multiple ARs can take place. In Ref. [33], $dI/dV$ at finite bias of an SNS Josephson junction exhibits conductance quantization in units of $\sim 2.7 \times 2e^2/h$, where the enhancement factor greater than 2 is a manifestation of multiple ARs. In addition, in SNS junctions quantized steps of the critical Josephson current emerge at zero bias as a result of the Josephson coupling through the quantized transport mode formed in the QPC.

In the SN junctions studied here, the absence of both multiple ARs and Josephson current allows us to study the behavior of conductance enhancement via a single AR near zero bias. Figure 2(b) shows the $V_{dc}$ dependence of $dI/dV$ measured on the first conductance plateau ($V_g = -1.2$ V). The data reveal a zero-bias peak with a half width at half maximum of 0.60 mV, in addition to small peaks at $|V_{dc}| \sim \Delta_0/e$ (= 1.28 mV) and oscillatory behavior at $|V_{dc}| > \Delta_0/e$. We emphasize that the observed zero-bias peak cannot be explained by the $\varepsilon$ dependence of the AR probability at the 2DEG/Nb interface alone. For an SN interface with a potential barrier, the Blonder-Tinkham-Klapwijk (BTK) model predicts that the AR probability has a maximum at $\varepsilon = \Delta_0$ but a minimum at $\varepsilon = 0$ [12] as shown in the inset of Fig. 2(b). Therefore, while the small peaks at $|V_{dc}| \sim \Delta_0/e$ can be understood as a manifestation of the maximum in the AR probability, the emergence of a zero-bias peak requires another mechanism that makes the conductance enhancement most efficient at $\varepsilon = 0$.

This observation suggests an analogy with reflectionless tunneling, which has been studied for disordered SN interfaces, i.e., a short mean free path in the N region and imperfect ARs at the SN interface [13,14,17,37]. In such SN junctions, frequent elastic scattering due to disorder in the N region allows normally reflected quasiparticles to be incident on the SN interface multiple times until they eventually undergo AR. Since the incident and Andreev-reflected quasiparticles share exactly the same dynamical phase at $\varepsilon = 0$, quantum interference between different paths is always constructive for $\varepsilon = 0$, which leads to the conductance enhancement [37]. On the other hand, an additional dynamical phase at $\varepsilon \neq 0$ randomizes the phase for different paths, resulting in the suppression of the conductance enhancement. In our hybrid QPCs, in which quasiparticle transport is ballistic, the role of disorder is played by the etched boundary of the 2DEG [38].

As we will show later in Fig. 4(a), the zero-bias peak is suppressed by a weak perpendicular magnetic field ($B_\perp \sim 7$ mT) much smaller than the critical field of Nb. The strong sensitivity to $B_\perp$ is consistent with the reflectionless tunneling model, in which a magnetic field of order $B_c = h/eA$—one magnetic flux quantum threading through the normal region with an area $A$—quenches the zero-bias peak [14]. In our case, $B_c$ is estimated to be $\sim 10$ mT [39], which is consistent with the experimental observation. Note that the cyclotron radius $r_c\ (= \hbar\sqrt{\pi n_s}/eB)$ under $B_\perp \sim 7$ mT is 16 μm, which is orders of magnitudes longer than $L_N$; this indicates that the orbital effect due to Lorentz force is negligible.

At biases greater than $\Delta_0/e$, $dI/dV$ oscillates with $V_{dc}$ [Fig. 2(b)]. These oscillations persist under in-plane magnetic fields greater than the critical field of Nb (data not shown). We therefore exclude the interference of Bogoliubov quasiparticles, known as the McMillan-Rowell oscillations [40], as the origin of the observed oscillations, and ascribe them to the Fabry-Pérot interference of electrons confined to the 2DEG cavity formed between the SN interface and QPC. The presence of Fabry-Pérot oscillations gives yet further evidence that the charge transport is ballistic and coherent in the cavity. For an electron with energy $\varepsilon$, the additional dynamical phase acquired during propagation through the cavity is given by $\Delta k \cdot 2L_N = [k(\varepsilon_{fN} + \varepsilon) - k(\varepsilon_{fN})] \cdot 2L_N \cong 2\varepsilon L_N/\hbar v_{fN}$ [41]. From the observed oscillation period of 3.5 mV [42], we obtain $v_{fN} = 3.7 \times 10^5$ m/s, which is considerably smaller than that calculated from $n_s$ ( $v_{fN} = \hbar\sqrt{2\pi n_s}/m^* = 9.3 \times 10^5$ m/s). This suggests that $n_s$ is reduced around the QPC and SN interface owing to etching-induced damage. It is also possible that the actual channel is longer than the designed length because of misalignment during fabrication.

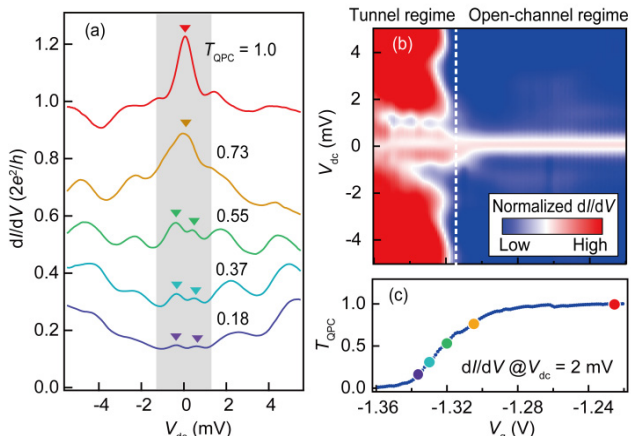

FIG. 3. (color online) (a) d$I$/d$V$ vs $V_{dc}$ at $T$ = 240 mK for several values of $V_g$. The arrows show peak positions. (b) 2D plot of d$I$/d$V$ as functions of $V_{dc}$ and $V_g$. To enhance the contrast of the peaks, the d$I$/d$V$ values are normalized by that at $V_{dc}$ = 0 mV. (c) $T_{QPC}$ vs $V_g$. $T_{QPC}$ is estimated from the d$I$/d$V$ at $V_{dc}$ = 2 mV on the basis of the Landauer formula. The dots show the $V_g$ values used for (a).

## IV. Spectroscopy of ABSs in the tunneling regime

Now we turn to the tunneling regime of the QPC to present results on the spectroscopy of ABSs, which are a hallmark of the phase coherent nature of quasiparticles. In this case, the QPC works both as a confining potential defining ABSs and a tunneling barrier for the spectroscopy. We show in Fig. 3(a) the d$I$/d$V$ spectra for several $V_g$'s in the single-channel regime. The corresponding values of $T_{QPC}$ are calculated from the d$I$/d$V$ in the linear-conductance regime using the Landauer formula d$I$/d$V$ = $2e^2/h \times T_{QPC}$. The calculated values are plotted as a function of $V_g$ in Fig. 3(c). In the calculation of $T_{QPC}$, we used the d$I$/d$V$ value at $V_{dc}$ = 2 mV ($> \Delta_0/e$) to avoid the influence of ARs on d$I$/d$V$. The conductance spectra in Fig. 3(a) exhibit distinct behavior for the high and low $T_{QPC}$. While a zero-bias peak is observed for $T_{QPC} > 0.6$, a double peak appears in the subgap regime $|V_{dc}| < \Delta_0/e$ for $T_{QPC} < 0.6$. This contrasting behavior is clearly seen in Fig. 3(b), where we plot the normalized differential conductance as a function of $V_g$ and $V_{dc}$. The plot also reveals that the position of the double peak ($V_{dc} = \pm 0.37$ meV) is nearly independent of $T_{QPC}$ below 0.6.

The model of de Gennes and Saint-James (dGSJ) [18] describes ABSs in a three-dimensional NINS structure. According to the model, for normal incident quasiparticles, the energy $\varepsilon_n$ of the $n$th ABS is given as the solution of the following equation:

$$\left(\frac{2}{\pi} \cdot \frac{L_N}{\xi_0}\right)\frac{\varepsilon_n}{\Delta_0} = n\pi + \arccos\left(\frac{\varepsilon_n}{\Delta_0}\right) \quad (n = 0, 1, \cdots).$$

The model predicts that only a single pair of ABSs is formed within $\Delta_0$ for $L_N/\xi_0 < 5.0$, and thus there is one solution for $L_N$ = 220 nm and $\xi_0$ = 152 nm ($L_N/\xi_0$ = 1.4), consistent with the experiment. However, the calculated ABS level is $|\varepsilon_0| \approx 0.76\Delta_0$ (= 0.97 meV), and this value is significantly higher than the position of the observed double peak (0.37 mV), suggesting the overestimation of $\xi_0$ (or $v_{fN}$). If we take the $v_{fN}$ value obtained from the Fabry-Pérot oscillations, we have $L_N/\xi_0$ = 3.6, which yields $|\varepsilon_0| \approx 0.47\Delta_0$ (= 0.60 meV), a better but not complete agreement with the experiment. A possible source of the disagreement is the finite probability of NR at the SN interface [43]: the presence of NRs lifts the degeneracy of ABSs and lowers the energy from that for the interface with perfect AR. Nevertheless, it is noteworthy that the simple dGSJ model captures the gross features of experimental observation in terms of the number and the energy position of ABSs, which supports the idea that the double peak originates from ABSs induced in the 2DEG.

We next turn our attention to the height and width of the conductance peaks in Fig. 3(a). Our experiment can be compared with the model of Riedel and Bagwell for a one-dimensional ballistic NINS structure [44]. The model predicts sharp peaks with the maximum conductance of $4e^2/h$ at the energies of the ABSs. The much lower peak height observed in our experiment reflects the fact that quasiparticles are confined in a two-dimensional conductor, in contrast to the one-dimensional NINS assumed in Ref. [44]. As we have discussed in the previous section, owing to the retro property of the AR, all the quasiparticles impinging on the SN interface at different incidence angles contribute to form ABSs, even after several NRs. Since the energy of an ABS depends on the travel distance (or the cavity length), the involvement of many different paths makes the peaks broader and smaller than expected for a one-dimensional structure.

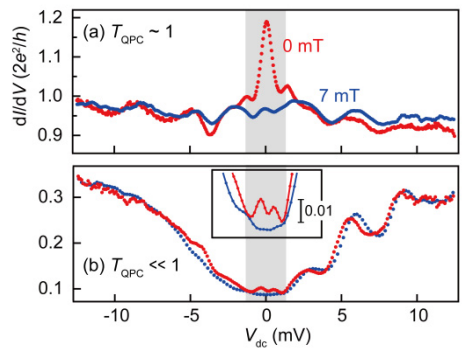

FIG. 4. (color online) Conductance spectra taken at $B_\perp$ = 0 (red) and 7 mT (blue) for (a) $T_{QPC} \approx 1$ and (b) $T_{QPC} \ll 1$. The inset in (b) shows a magnified view around $V_{dc}$ = 0 mV.

Finally, we examine the effects of a perpendicular magnetic field $B_\perp$ on the conductance spectra. Since ARs require time-reversal symmetry, AR-related phenomena are expected to be susceptible to an external magnetic field. We indeed observe that $B_\perp$ as small as 7 mT suppresses the AR-induced zero-bias peak for $T_{QPC} \sim 1$ [Fig. 4(a)]. Moreover, the same magnetic field eliminates the double peak for $T_{QPC} <$ 0.6 [Fig. 4(b)], providing further evidence that ABSs are responsible for the observed double peak.

## V. Conclusion

Using a QPC as a mode-selective tunable-transmission probe, we have observed two experimental signatures revealing the coherent nature of Bogoliubov quasiparticles in $In_{0.75}Ga_{0.25}As$ 2DEG coupled to a Nb electrode. Firstly, in the open-channel regime, the observation of a zero-bias peak with single-channel conductance exceeding $2e^2/h$ demonstrates the transmission of Andreev-reflected holes through the QPC. The bias and magnetic field dependences of the zero-bias peak suggest a mechanism analogous to reflectionless tunneling, indicating the coherent nature of quasiparticle transport. Secondly, tunneling spectroscopy using the QPC in the tunneling regime clearly probes the formation of ABSs in 2DEG-based SN junctions in the ballistic regime, an observation that has not been previously reported. Our results thus encourage future studies on more complex 2DEG-based hybrid SN structures integrating low-dimensional structures such as nanowires and quantum dots defined by electrostatic gating. Such systems would allow for the manipulation of ABSs, a necessary step toward novel electronics that can exploit AR.

**Acknowledgements**

This work was supported by Japan Society for the Promotion of Science (JSPS) KAKENHI (Grant Nos. 22103002 and 15H05854).

# Andreev reflection and bound state formation in a ballistic two-dimensional electron gas probed by a quantum point contact

Supplementary material

## DETERMINATION OF THE FABRY-PÉROT OSCILLATION PERIOD

Figure S1(a) shows a contour plot of $dI/dV$ as a function of $V_g$ and $V_{dc}$, where the Fabry-Pérot oscillation manifests as the stripe features superimposed on the QPC's nonlinear conductance. The peak positions of the oscillation are almost equally spaced with respect to the peak index as shown in Fig. S1(b). To determine the oscillation period that is used for calculating $v_{fN}$ in the main text, we first extracted oscillation periods at multiple $V_g$'s in the single-channel regime ($V_g$ = -1.17, -1.26, -1.35 V) and averaged them.

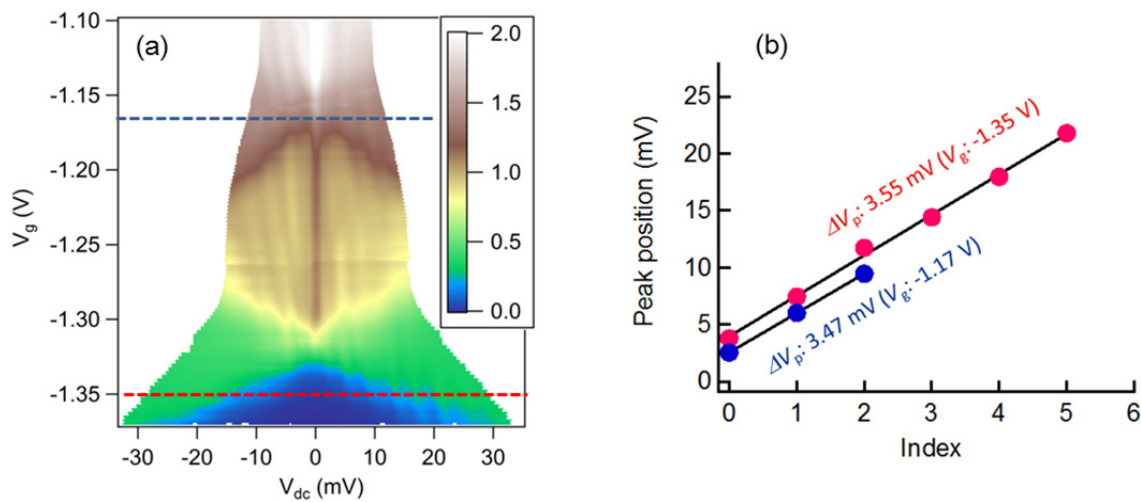

FIG. S1. (a) Contour plot of $dI/dV$ as a function of $V_g$ and $V_{dc}$. (b) Peak positions of the Fabry-Pérot oscillation at typical $V_g$ values of -1.17 and -1.35 V, which are also represented by horizontal dashed lines in (a).